\documentclass[aps,preprint,superscriptaddress,longbibliography,prl]{revtex4-1}
\usepackage{graphicx, amsmath, verbatim, dsfont, amsfonts, color, amssymb}
\usepackage{braket}
\usepackage[us]{datetime}

\begin{document}

\title{Transport in helical Luttinger liquids in the fractional quantum Hall regime}

\author{Ying~Wang}
\affiliation{Department of Physics and Astronomy, Purdue University, West Lafayette, IN 47907 USA}
\affiliation{Birck Nanotechnology Center, Purdue University, West Lafayette, IN 47907 USA}
\author{Vadim Ponomarenko}
\affiliation{Department of Physics and Astronomy, Purdue University, West Lafayette, IN 47907 USA}
\affiliation{Ioffe Physico-Technical Institute,  194021,  Saint-Petersburg, Russia}
\author{Kenneth W. West}
\affiliation{Department of Electrical Engineering, Princeton University, Princeton, NJ 08540 USA}
\author{Kirk Baldwin}
\affiliation{Department of Electrical Engineering, Princeton University, Princeton, NJ 08540 USA}
\author{Loren N. Pfeiffer}
\affiliation{Department of Electrical Engineering, Princeton University, Princeton, NJ 08540 USA}
\author{Yuli Lyanda-Geller}
\email{yuli@purdue.edu}
\affiliation{Department of Physics and Astronomy, Purdue University, West Lafayette, IN 47907 USA}
\affiliation{Birck Nanotechnology Center, Purdue University, West Lafayette, IN 47907 USA}
\author{Leonid~P.~Rokhinson}
\email{leonid@purdue.edu}
\affiliation{Department of Physics and Astronomy, Purdue University, West Lafayette, IN 47907 USA}
\affiliation{Birck Nanotechnology Center, Purdue University, West Lafayette, IN 47907 USA}
\affiliation{Department of Electrical and Computer Engineering, Purdue University, West Lafayette, IN 47907 USA}

\date{December 31, 2020} 

\maketitle

\section{abstract} 

Domain walls in fractional quantum Hall ferromagnets are gapless helical one-dimensional channels formed at the boundaries of topologically distinct quantum Hall (QH) liquids. Na\"{i}vely, these helical domain walls (hDWs) constitute two counter-propagating chiral states with opposite spins. Coupled to an s-wave superconductor, helical channels are expected to lead to topological superconductivity with high order non-Abelian excitations \cite{Lindner2012,Clarke2012,Alicea2016}. Here we investigate transport properties of hDWs in the $\nu=2/3$ fractional QH regime. Experimentally we found that current carried by hDWs is substantially smaller than the prediction of the na\"{i}ve model. Luttinger liquid theory of the system reveals redistribution of currents between quasiparticle charge, spin and neutral modes, and predicts the reduction of the hDW current. Inclusion of spin-non-conserving tunneling processes reconciles theory with experiment. The theory confirms emergence of spin modes required for the formation of fractional topological superconductivity.

\section{introduction} 

Gapless chiral edge states, a hallmark of the quantum Hall effect (QHE), are formed at the boundaries of the two-dimensional (2D) electron liquid. These states are protected due to their topological properties; their spatial separation suppresses backscattering and insures precise quantization of the Hall conductance over macroscopic distances \cite{PerspQHE2007}. Symmetry-protected topological systems can support spatially coexisting counter-propagating states. For example, in 2D topological insulators time reversal symmetry insures orthogonality of Kramers doublets \cite{Ando2002,Xu2006,Bardarson2008}; in graphene the conservation of angular momentum prevents backscattering  in the quantum spin Hall effect regime \cite{Young2014}. Local symmetry protection is not as robust as spatial separation in the QHE and, as a result, hDWs have finite scattering and localization lengths. Helical states can be also engineered by arranging proximity of two counter-propagating chiral states with opposite polarization, e.g. in electron-hole bi-layers \cite{Sanchez-Yamagishi2017} or double quantum well structures \cite{Ronen2018}, where local charge redistribution between two quantum wells creates two counter-propagating chiral states at the boundary of quantum Hall liquids with different filling factors. In the latter system, spatial separation into two quantum wells suppresses the inter-channel scattering, and transport in each chiral channel is found to be ballistic over macroscopic distances.

An intriguing possibility to form helical channels in the interior of a 2D electron gas is to induce a local quantum Hall ferromagnetic transition. In the integer QHE regime, scattering between overlapping chiral edges from different Landau levels is suppressed due to the orthogonality of the wavefunctions, but spin-orbit interaction mixes states with opposite spins and opens a small gap in the helical spectrum \cite{Kazakov2016,Kazakov2017,Simion2018}. In the fractional quantum Hall effect (FQHE) regime, electrostatically-controlled transition between unpolarized (\textit{u}) and polarized (\textit{p}) $\nu=2/3$ states results in the formation of a conducting channel at the boundary between \textit{u} and \textit{p} regions (filling factor $\nu^{-1}=B/n\phi_0$, where $B$ is an external magnetic field, $\phi_0=h/e$ is a flux quanta and $n$ is electron density). Superficially, transition between \textit{u} and \textit{p} states in the bulk can be understood as a crossing of two composite fermion energy states with opposite spin polarization \cite{Ronen2018,Jain1989,Kukushkin1999}. Within this model, the hDW at the \textit{u-p} boundary consists of two counter-propagating chiral states with opposite spin and fractionalized charge excitations, and presents an ideal platform to build fractional topological superconductors with parafermionic and Fibonacci excitations \cite{Lindner2012,Clarke2012,Vaezi2013,Mong2014,Vaezi2014,Liang2019}.
Highly correlated $\nu=2/3$ state exhibits rich physics beyond an oversimplified model of $\nu^*=2$ integer QHE for weakly interacting composite fermions and includes observation of upstream neutral modes \cite{Gurman2012,Inoue2014,Rosenblatt2017,Wang2013}, short-range upstream charge modes \cite{Lafont2019}, and a crossover from $e^*=1/3$ to $e^*=2/3$ charge excitations in shot noise measurements \cite{Bid2009}. In the bulk, the spin transition at $\nu=2/3$ is accompanied by nuclear polarization \cite{Kronmueller1998,Kraus2002,Stern2004} indicating spin-flip processes in the 2D gas, a phenomena not seen in bi-layer systems \cite{Ronen2018,Cohen2019}. Thus, we expect a hDW  formed at a boundary between \textit{u} and \textit{p} $\nu=2/3$ states to be more complex than a simple overlap of two non-interacting chiral modes.

Here we study the electron transport in samples where hDWs of different length $L$ are formed by electrostatic gating. Experimentally, in the limit $L\rightarrow0$ approximately $12\%$ of the edge current is diverted into the hDW, a number drastically different from the 50\% prediction for two non-interacting counter-propagating chiral channels. To address this discrepancy theoretically we consider tunneling between Luttinger liquid modes  \cite{Wen1995} through a hDW in the strong coupling limit \cite{Sandler1999,Nayak1999,Moore2002} . We found that in the presence of a strong inter-edge tunneling edge channels in \textit{u} and \textit{p} regions populate unequally, both at the boundary of the 2D gas and within the hDW, forming a number of down- and up-stream charge, spin and neutral modes.  For spin-conserving tunneling 1/4 of the incoming charge current is diverted into the hDW, while allowing spin-flip processes further reduces hDW current. Indeed, at high bias currents we observe an increase in the current carried by the hDW. This indicates formation of a bottleneck for spin flips due to Overhauser pumping of nuclei and a crossover from spin-non-conserving to spin-conserving transport.

\begin{figure}
\def\ffile{f-exp}
\centering
\includegraphics[width=0.98\textwidth]{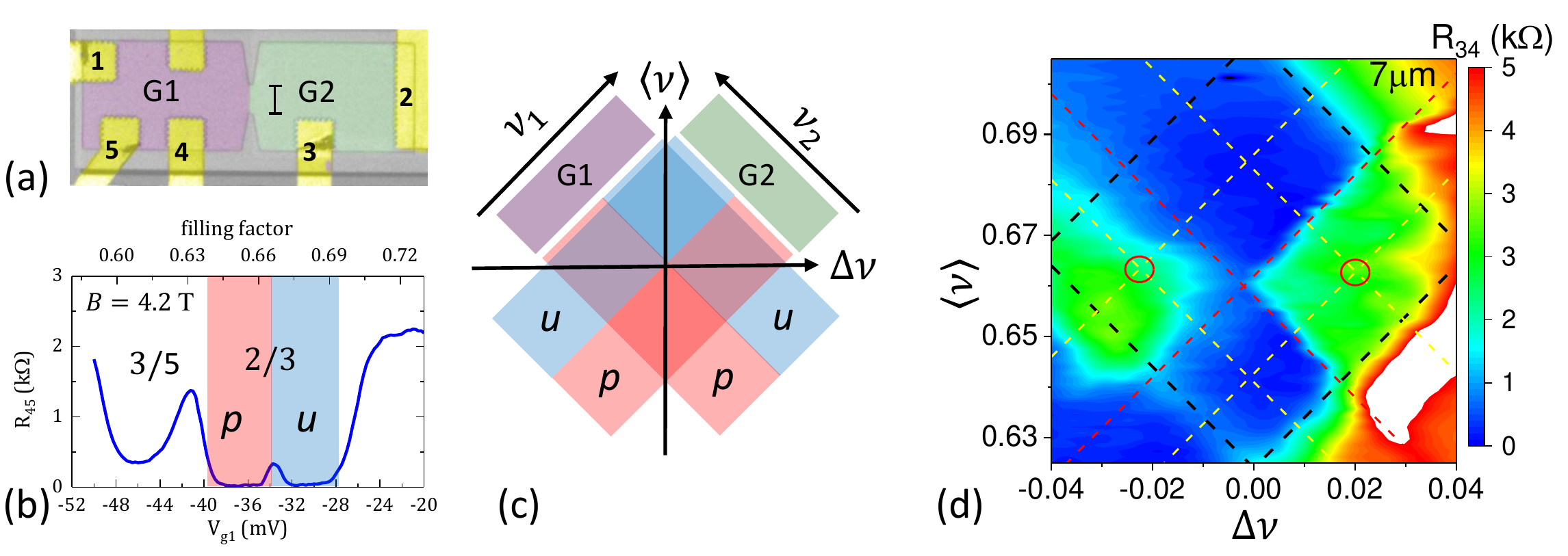}
\caption{\textbf{Formation of helical domain walls at $\nu=2/3$.} (a) A false color image of a typical device. Yellow regions are ohmic contacts, 2D gas in green and magenta regions is controlled by gates G1 and G2, in the grey area 2D gas is removed. The bar is 7 $\mu$m. (b) Magnetoresistance $R_{45}=(V_5-V_4)/I$ of a 2D gas is plotted as a function of gate voltage (controlling filling factor $\nu$) at a fixed $B=4.2$ T. Small peak at $-34$ mV is the phase transition between unpolarized ({\it u}) and polarized ({\it p}) $\nu=2/3$ FQHE liquids. (c) A diagram of  {\it u} and {\it p} states as a function of $\nu_1$ and $\nu_2$ under gates $G1$ and $G2$; $\langle\nu\rangle=(\nu_1+\nu_2)/2$ and $\Delta\nu=(\nu_1-\nu_2)$.  (d) Resistance $R_{34}=(V_4-V_3)/I$ across a $7\ \mu$m - long gates boundary is plotted as a function of $\langle\nu\rangle$ and $\Delta\nu$. Black square outlines the $\nu=2/3$ region, red lines mark \textit{u-p} transitions and yellow lines mark centers of \textit{u} and \textit{p} regions.}
\label{\ffile}
\end{figure}

\section{Experimental Results} 

Several devices in a Hall bar geometry with multiple gates have been fabricated in order to study transport through hDWs, Fig.~\ref{f-exp} (for heterostructure and fabrication details see Methods). Devices are separated into two regions $G1$ and $G2$; electron density $n$ in these regions can be controlled independently by electrostatic gates. In the IQHE regime when filling factors $\nu$ under gates $G1$ and $G2$ differ by one, a single chiral channel is formed along the gates boundary and resistance $R_{34}$ measured across the boundary is either quantized or zero depending on the sign of the filling factor gradient and direction of $B$ (see Supplementary Material). Likewise, a chiral channel is formed when gates boundary separates two different FQHE states.

Fractional QHE states can be understood as integer QHE states for composite fermions in a reduced field, $\nu=\nu^*/(2\nu^*\pm 1)$, where $\nu^*$ is the filling factor for composite fermions \cite{JainCFbook2007}. Energy separation between composite fermions levels depends on the competition between charging energy $E_c\propto \sqrt{B}$ and Zeeman energy $E_Z\propto B$, and the two lowest energy levels with opposite spins 0-down and 1-up cross at finite $B^*>0$ due to different field dependencies. When level crossing occurs within the $\nu=2/3$ plateau ($\nu^*=2$ for composite fermions), the energy gap for quasiparticle excitations vanishes providing mechanism for charge backscattering and, hence, at $B=B^*$ resistance of the 2D gas is no longer zero. In our devices it is possible to control $B^*$ by electrostatic gating, and a small peak within the 2/3 plateau in Fig.~\ref{f-exp}b is a quantum Hall ferromagnetic transition between polarized (\textit{p}) and unpolarized (\textit{u}) regions.

Independent control of filling factors $\nu_1$ and $\nu_2$ under G1 and G2 divides the 2/3 region into four quadrants \textit{uu}, \textit{pp}, \textit{up} and \textit{pu}, where the  first letter corresponds to polarization of the state under G1 and the second corresponds to polarization under G2.
Within the Landauer-B\"uttiker formalism \cite{Beenakker1990,Brey1994} resistance $R_{34}=(1/\nu_1-1/\nu_2)R_q$ should be zero for all combinations of polarizations under the gates since quantum numbers $\nu_1=\nu_2=2/3$ for both \textit{u} and \textit{p} states (here $R_q$ is the Klitzing's constant). Experimentally $R_{34}$ is found to be vanishingly small in \textit{uu} and \textit{pp} quadrants, consistent with a single topological state being extended over the whole device. $R_{34}>0$ in \textit{up} and \textit{pu} quadrants indicates backscattering between edge channels and, which, combined with zero longitudinal resistance under G1 and G2, means formation of a conducting channel along the gates boundary. Unlike resistance measured across chiral channels formed, e.g., between $\nu=2/3$ and $\nu=3/5$ FQHE states (see Supplementary Material), resistance measured across the boundary of \textit{u} and \textit{p} quantum liquids at $\nu=2/3$ shows almost no dependence on the direction of the external magnetic field and density gradient, consistent with the formation of a helical domain wall \cite{Wu2018}.

\begin{figure}[t]
\def\ffile{f-idw}
\centering
\includegraphics[width=0.8\textwidth]{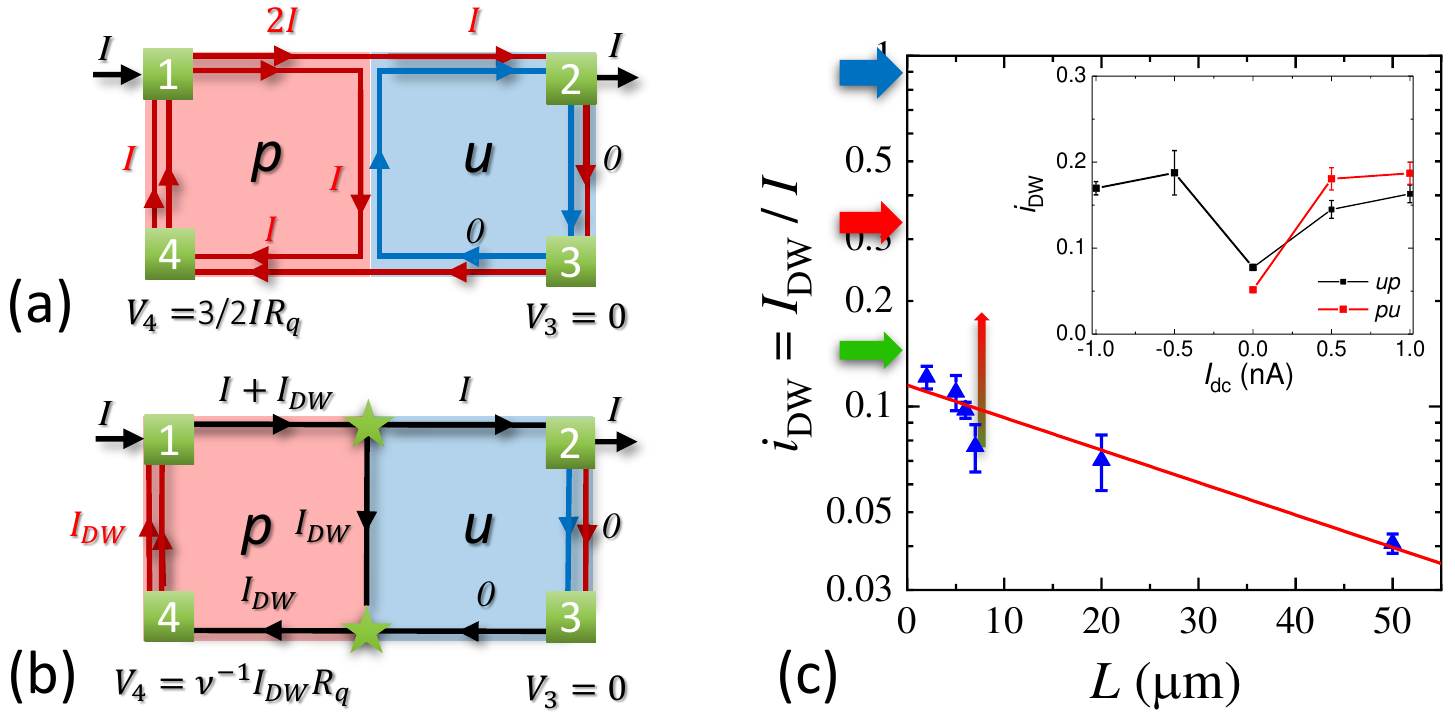}
\caption{\textbf{Conduction of helical domain walls.} (a) A simplified picture of non-interacting chiral edge modes at $\nu=2/3$. An inner spin-up edge (red) in {\it p} state carries current $I$ while a spin-down edge (blue) in {\it u} state carries no current. (b) Charge conservation and chirality of edge states set the potential $V_4=\nu^{-1}I_{DW}R_q$  to be proportional to the current $I_{DW}$ diverted via the helical domain wall. (c) Scaling of the domain wall current $i_{DW}=I_{DW}/I$ with hDWs length $L$. The values are averaged between \textit{up} and \textit{pu} states (red circles in Fig.~\ref{f-exp}d). Red line is a fit to an exponential decay with the $L=0$ value $i_{DW}^0=0.115$ and the decay length $L_0=47\ \mu$m.  Arrows indicate $i_{DW}^0$ values expected for non-interacting edge model (blue) and Luttenger liquid model with (green) and without (red) spin flips. In the inset dependence of $i_{DW}$ on large external dc current is plotted for $7\ \mu$m hDW for \textit{up} and \textit{pu} gates configuration.}
\label{\ffile}
\end{figure}

Protection of helical states from backscattering and localization is weaker than for spatially separated chiral edge states, and conduction of hDWs is length dependent. In Fig.~\ref{f-idw} a fraction of the external current $I$ that flows through the hDW, $i_{DW}=I_{DW}/I$, is plotted as a function of hDW length $L$. $i_{DW}$ is found to decrease exponentially with $L$, $i_{DW}=i_{DW}^0\exp[-L/L_0]$, with a characteristic length $L_0=47\ \mu$m. The value of $i_{DW}^0$ corresponds to the transport through a ballistic hDW in the absence of localization. Within a simplified model of $\nu=2/3$ edge states consisting of equally populated $1/3+1/3$ chiral modes and no interaction between chiral channels with opposite spin polarization (Fig.~\ref{f-idw}a), one expects $i_{DW}^0=1$ (marked by a blue arrow in Fig.~\ref{f-idw}c), an order of magnitude larger than the experimentally observed value (marked by a green arrow). Note that transport through helical modes formed in double quantum well structures are well described by this simple model of weakly interacting chiral states \cite{Ronen2018}.

\section{Theory} 

An isolated hDW at a boundary of \textit{p} and \textit{u} phases was studied in Refs.~\onlinecite{Wu2018,Liang2019}, where disk and torus geometry were employed to avoid physical edges of the sample and coupling of domain wall modes to these edges. Analytical model and numerical results indicate an existence of modes with opposite velocities and spins within the hDW region, a prerequisite for generating topological superconductivity. No neutral or spin modes appear within the K-matrix Luttinger liquid approach \cite{Wen1995} in these isolated hDW models. Inclusion of a superconducting pairing term into the hDW Hamiltonian leads to the 6-fold degeneracy of the ground state within a range of parameters, a signature of low energy parafermionic excitations.

To calculate scattering of edge modes at a sample boundary by a hDW we need to move beyond an isolated hDW model. Here we consider tunneling of fractional QH edge states through the hDW in the strong coupling limit. The presence of the hDW imposes boundary conditions on the incoming and propagating modes and affects the population of charge and spin modes along the edges of the \textit{u} phase and charge and neutral modes along the edges of the \textit{p} phase. In terms of the separated charged and neutral modes $\varphi_c$ and $\varphi_n$, the Luttinger liquid action for $\nu =2/3$ hierarchical edge states for the \textit{p} phase reads
\begin{equation}
S = -\frac{1}{4\pi}\int dt \int dx \left[ -3\partial_x \varphi_c \left(\partial_t+ v_c\partial_x \right)\varphi_c   +  \varphi_n\left(\partial_t- v_n\partial_x \right)\varphi_n\right],
\end{equation}
where $v_c$ and $v_n$ are velocities of charge and neutral modes, correspondingly. The neutral mode density measures the difference in the occupation of the first two composite fermion $\Lambda$-levels. The \textit{u} phase is described by a similar action, in which the neutral $n$-mode becomes the spin $s$-mode that determines a spin current in the \textit{u} phase.

In the presence of voltage $V$, change in the mode $\varphi_c$ due to charge injection is characterized by an average induced current
${\bar j}= \frac{e^2V}{3\pi\hbar}$.    Tunneling between \textit{p} and \textit{u} phases carried by electrons with the same spin can be described in terms of quasiparticles by the tunnel Hamiltonian
\begin{equation}
{\cal H}_T=-{\tilde t}\cos{\left(\frac{1}{\sqrt{2}}\left[3\varphi_{cp}-\varphi_{np}-3\varphi_{cu}+\varphi_{su} \right] \right)}.
\label{eq:Ht}
\end{equation}

\begin{figure}[t]
\centering
\def\ffile{f-chs}
\includegraphics[width=0.9\textwidth]{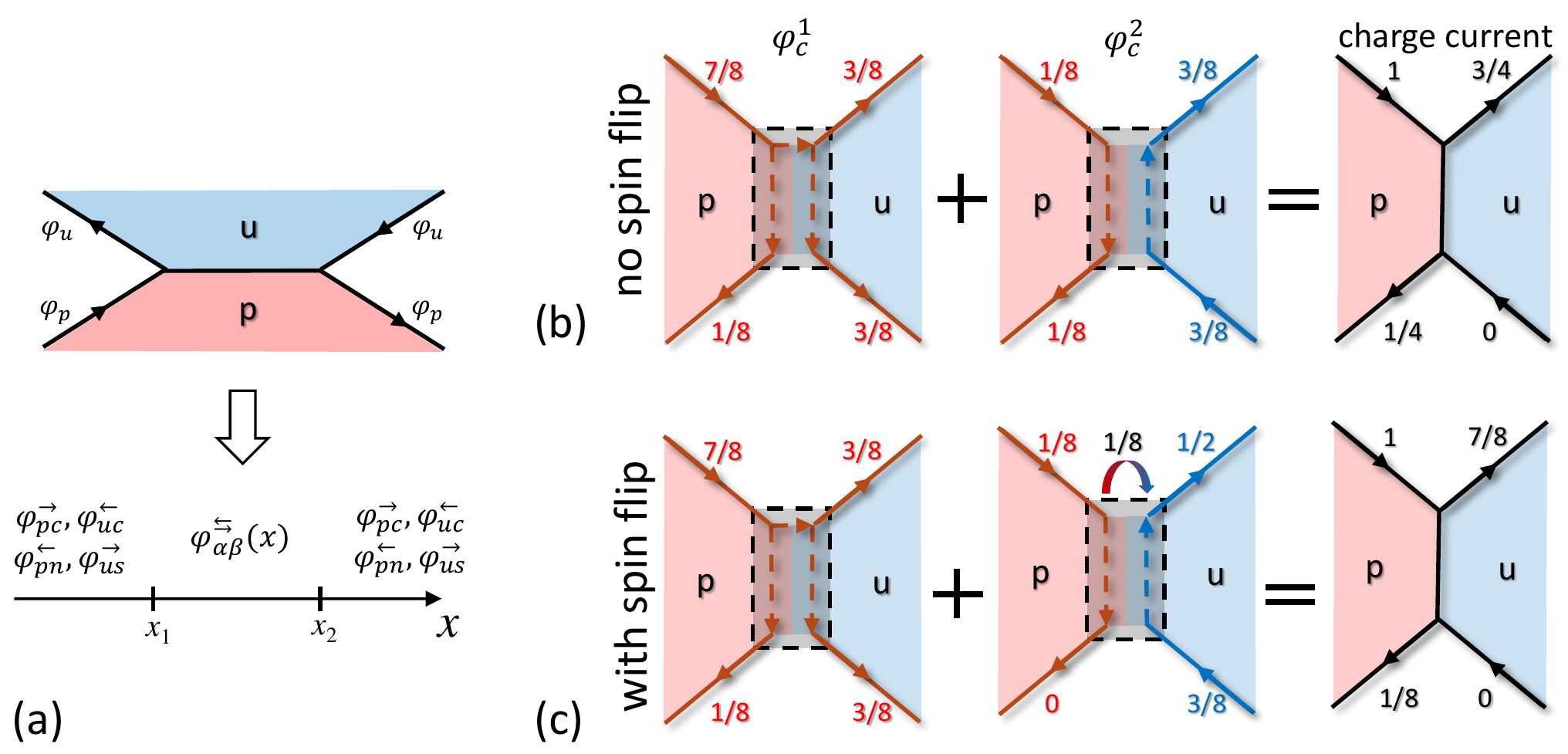}
\caption{\textbf{Schematic representation of bosonic modes.} (a) Mapping of bosonic modes $\varphi_{\alpha}$ along sample edges onto a 1D Luttinger model modes  $\varphi_{\alpha\beta}^{\rightleftarrows}(x)$ for a domain wall with length $L=x_2-x_1$. Subscripts $\alpha=\{p,u\}$ label polarized and unpolarized phases, $\beta=\{c,n,s\}$ is for charge, neutral and spin modes and an arrow in the superscript specifies projection of the mode's group velocity. Arrows on the edges define chirality of edge channels. (b,c) Visualization of bosonic charge modes $\varphi^1_c$ and $\varphi^2_c$ propagation without (b) and with (c) spin flips. Red (blue) mode color indicates a spin-up (spin-down) polarization. Numbers indicate the fraction of the incoming current carried by the mode.}
\label{\ffile}
\end{figure}

Mapping of an edge-hDW-edge structure onto one dimensional bosonic modes $\varphi(x)$ is shown schematically in Fig.~\ref{f-chs}a. In general, hDW and each region outside of the hDW may contain up to eight bosonic modes $\varphi_{\alpha\beta}^{\rightleftarrows}(x)$, where $\alpha=\{p,u\}$, $\beta=\{c,n/s\}$ and the superscript \{$\leftarrow,\rightarrow$\} indicates projection of the velocity $v_{\beta}$ on the $x$-axis. However, chirality of edge channels reduces the number of bosonic modes to four. Therefore, it is convenient to consider two charge modes $\varphi_{pc}^{\rightarrow}$ and $\varphi_{uc}^{\leftarrow}$ and two spin/neutral modes $\varphi_{pu}^{\leftarrow}$ and $\varphi_{us}^{\rightarrow}$.

In the strong coupling limit ${\tilde t}\rightarrow \infty$  charge, neutral and spin currents can be found by imposing the following boundary conditions on bosonic fields right outside of the hDW $[x_1,x_2]$:
\begin{equation}
\begin{pmatrix}
\varphi^{\rightarrow}_{pc}(x_2+0)\\[\jot]
\varphi^{\rightarrow}_{us}(x_2+0)\\[\jot]
\varphi^{\leftarrow}_{uc}(x_1-0)\\[\jot]
\varphi^{\leftarrow}_{pn}(x_1-0)
\end{pmatrix}
=\frac{1}{4}\left(\begin{array}{cccc}1 & -1&3&1  \\-3 &3&3&1\\
3&1&1&-1\\3&1&-3&3\end{array}\right)
\begin{pmatrix}
\varphi^{\rightarrow}_{pc}(x_1-0)\\[\jot]
\varphi^{\rightarrow}_{us}(x_1-0)\\[\jot]
\varphi^{\leftarrow}_{uc}(x_2+0)\\[\jot]
\varphi^{\leftarrow}_{pn}(x_2+0)
\end{pmatrix},
\label{BC}
\end{equation}
which connect all incoming modes at the right side and outgoing modes at the left side of the equation.
Details of calculation, including the account of scattering between \textit{p}-modes in the polarized region, is presented in Supplementary Materials. The solution for charge fields $\varphi_c$ (Eq.~\ref{BC}) is shown schematically in Fig.~\ref{f-chs}(b,c) where we assume a unit current incident on the hDW, this current corresponds to $(I+I_{DW})$ in Fig.~\ref{f-idw}b. There are two quasiparticle bosonic modes along the edges on both \textit{p} and \textit{u} sides which we refer to as $\varphi^1_c$ and $\varphi^2_c$. An incoming charge current is divided between $\varphi^1_c$ and $\varphi^2_c$ in a ratio 7:1. 1/7 of the $\varphi^1_c$ mode is scattered back into the \textit{p} phase across the hDW, while 6/7 is tunneling into the \textit{u} phase and splits equally between a downstream and an upstream (along the hDW) modes. The $\varphi^2_c$ bosonic mode has different spin polarization on \textit{p} and \textit{u} sides, it carries downstream 1/8 of the total incoming current in \textit{p} and 3/8 in \textit{u} phases. In the absence of spin flips there is no tunneling between $\varphi^2_{cp}$ and $\varphi^2_{cu}$ modes. The total current backscattered through the hDW $I_{DW}=1/4(I+I_{DW})$ or $i_{DW}=1/3$. This value is three times larger than the experimentally measured $i_{DW}$ and is indicated by a red arrow in Fig.~\ref{f-idw}c.

In 2D gases formed in GaAs heterostructures spin transition at $\nu=2/3$ is accompanied by a dynamic nuclear spin polarization \cite{Li2008book}. Its main mechanism is the hyperfine coupling of electron and nuclear spins, which for QHE plateaus is usually suppressed due to a large difference between electron and nuclear Zeeman splittings. However, near the \textit{u}-\textit{p} phase transition electronic states with spin-up and spin-down are almost degenerate, enabling hyperfine coupling. This spin-flip mechanism can couple $\varphi^2_c$-up and $\varphi^2_c$-down bosonic modes on two sides of the hDW, as shown schematically in Fig.~\ref{f-chs}c. Assuming this charge transfer induces forward scattering and reduces the fraction of the total current flowing in the hDW from 1/4 to 1/8, which corresponds to $i_{DW}=1/7$ (green arrow in Fig.~\ref{f-idw}c), we  obtain the value in a good agreement with the experimentally observed $i_{DW}^0\approx 0.11$.

To test the role of spin flips it is possible to pass a large dc current and polarize nuclei in the vicinity of the tri-junction. Saturation of nuclear spin polarization is expected to create a bottleneck for electron spin flips and disable charge transfer between two $\varphi_c^2$ bosonic modes with opposite polarization. Indeed, application of $I_{dc}>0.5$ nA results in approximately 3-fold increase of $i_{DW}$, as shown in the inset in Fig.~\ref{f-idw}. A corresponding shift of $i_{DW}$ for the 7 $\mu$m hDW is shown with a vertical arrow on the main plot. This shift is consistent with the 2.3 times current increase expected for the crossover from spin-flip-dominated to no-spin-flip transport.

\section{Methods} 

Devices are fabricated from GaAs/AlGaAs inverted single heterojunction heterostructures
with electron gas density $0.9\cdot 10^{11}$ cm$^{-2}$ and mobility $5\cdot 10^{6}$ cm$^{2}/Vs$. Details of heterostructure design can be found in \onlinecite{Wan2015}, these heterostructures demonstrate efficient electrostatic control of the spin transition at $\nu=2/3$ \cite{Wu2018}. Devices are patterned in a Hall bar geometry using e-beam lithography and wet etching, photograph of a typical sample is shown in Fig.~\ref{f-exp}. Devices are divided into two regions by semi-transparent gates (10 nm of Ti), the gates are separated from the surface of the wafer and from each other by 50nm of Al$_2$O$_3$ grown by the atomic layer deposition (ALD). Gates boundary is aligned with a $2-50$ $\mu$m - wide mesa constriction. Ohmic contacts are formed by annealing Ni/Ge/Au 30nm/50nm/100nm at 450$^\circ$C for 450 seconds in $H_2/N_2$ atmosphere. Measurements are performed in a dilution refrigerator at a base temperature of 20 mK using conventional low frequency lock-in technique with excitation current $I_{ac}=100$ pA. A 2D electron gas is formed by shining red LED at 4 K.

\textbf{Acknowledgments}\\
Experimental part of the work is supported by NSF award DMR-1836758 (Y.W. and L.P.R.).  Theoretical work is supported by the U.S. Department of Energy, Office of Basic Energy Sciences, Division of Materials Sciences and Engineering under Award DE-SC0010544 (Y.L-G). Heterostructures development and growth is funded in part by the Gordon and Betty Moore Foundation’s EPiQS Initiative, Grant GBMF9615 to L.N.P, and by the NSF MRSEC grant DMR-1420541.

\textbf{Author contributions}\\
Y.W. L.P.R conceived and performed experiments, V.P. and Y.L.G developed theory, K.W.W, K.B and L.N.P. developed and grew heterostructure materials. Y.W, Y.L.G and L.P.R have written the manuscript.



\clearpage
\newpage

\renewcommand{\thefigure}{S\arabic{figure}}
\renewcommand{\theequation}{S\arabic{equation}}
\renewcommand{\thepage}{sup-\arabic{page}}
\setcounter{page}{1}
\setcounter{equation}{0}
\setcounter{figure}{0}

\begin{center}
\textbf{\large Transport in helical Luttinger liquids in the fractional quantum Hall regime}\\
Ying~Wang, Vadim Ponomarenko, Kenneth W. West, Kirk Baldwin, Loren N. Pfeiffer, Yuli Lyanda-Geller, Leonid~P.~Rokhinson\\
\textbf{Supplementary Materials}
\end{center}

\section{Extended Figures}

\begin{figure}[h]
\def\ffile{fs-Rext}
\centering
\includegraphics[width=0.98\textwidth]{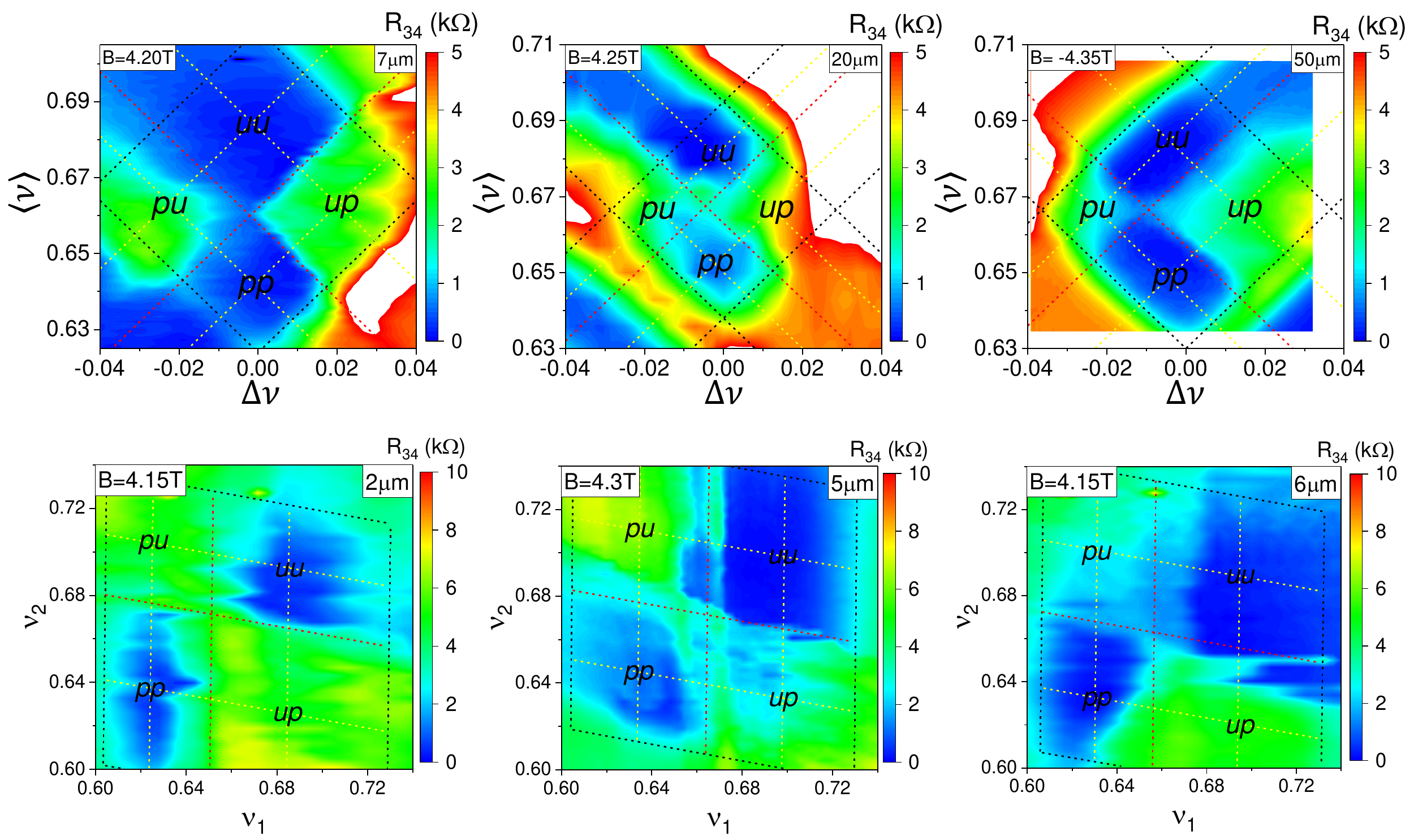}
\caption{Resistance $R_{34}$ is plotted as a function of $\langle\nu\rangle$ and $\Delta\nu$ for samples with $7\ \mu$m, $20\ \mu$m and $50\ \mu$m gates boundary and as a function of $\nu_1$ and $\nu_2$ for samples with $2\ \mu$m, $5\ \mu$m and $6\ \mu$m gates boundary. Black lines outlines the $\nu=2/3$ region, red lines mark \textit{u-p} transitions and yellow lines mark centers of \textit{u} and \textit{p} regions.}
\label{f-Rext}
\end{figure}

\newpage
\section{Transport in the presence of a chiral channel}

\begin{figure}[h]
\def\ffile{fs-chiral}
\centering
\includegraphics[width=0.9\textwidth]{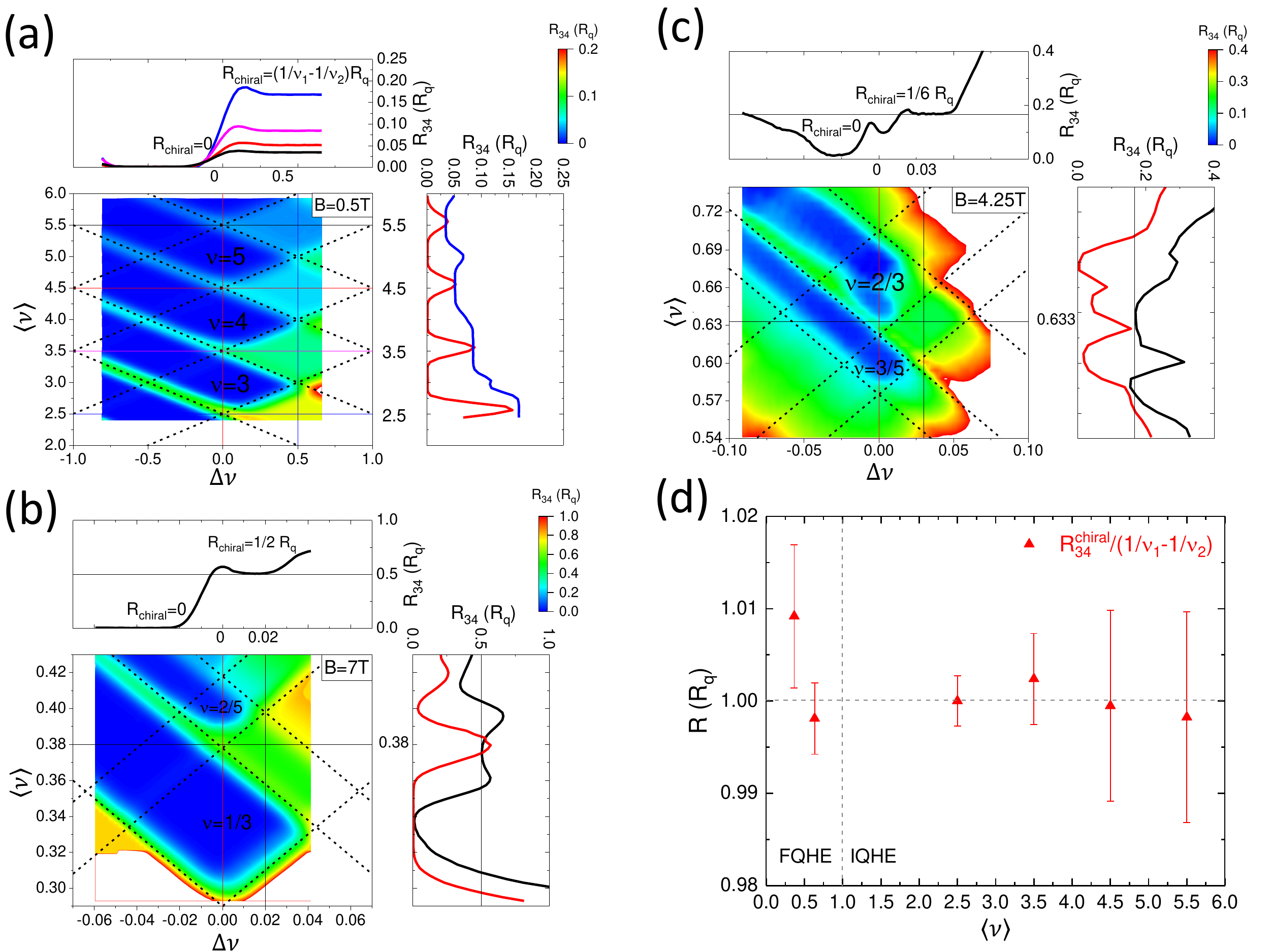}
\caption{Chiral channels in IQHE and FQHE regimes: (a-c) Resistance $R_{34}$ is plotted as a function of $\langle\nu\rangle$ and $\Delta\nu$ for the sample with  $20\ \mu$m gates boundary. The top and right panels in each plot show line cuts indicated in color plots by solid lines. Resistance $R_{34}$ is quantized for $\Delta\nu=+1$ in the IQHE regime and for $\Delta\nu^*=+1$ in the FQHE regime ($\nu^*$ is a composite fermion filling factor), $R_{34}=0$ for $\Delta\nu=-1$ and $\Delta\nu^*=-1$. (d) $R_{34}/(1/\nu_1-1/\nu_2)$ for chiral channels is plotted as a function of a filling factor.}
\label{f-chiral}
\end{figure}

In order to contrast helical and chiral channels we perform measurements of $R_{34}$ when filling factors $\nu_1$ and $\nu_2$ under gates $G1$ and $G2$ are different quantum Hall liquids. According to Landauer-B\"uttiker theory \cite{Beenakker1990,Brey1994} for $B>0$ $R_{34}=(\frac{1}{\nu_1}-\frac{1}{\nu_2})\cdot R_q$ for $\Delta\nu>0$ and $R_{34}=0$ for $\Delta\nu<0$ (zero and non-zero values will be switched for $B<0$).
In Figure \ref{f-chiral}abc, resistance $R_{34}$ is plotted over a range of filling factors in IQHE and FQHE regimes. In (d) we plot experimentally measured $R_{34}$ scaled by $(\frac{1}{\nu_1}-\frac{1}{\nu_2})$; the values fall within 1\% of the expected values.

\newpage
\section{Modeling of transport through a helical domain wall at $\nu=2/3$}

\subsection{Description of edges of $\nu=2/3$ state in terms of bosonic fields and quasiparticle bosonic fields}

Luttinger liquid action for $ \nu=2/3$  hierarchical edge states in terms of bosonic fields of \textbf{\textit{electron operators}} $\Phi_{1,2}$
\begin{equation}
S = -\frac{1}{4\pi}\int dt \int dx \left[ \partial_x \Phi \hat{K}^{-1}\partial_t\Phi   +  \partial_x \Phi\hat{V}^{e}  \partial_x \Phi\right],
\end{equation}
and the charge density is
\begin{equation}
\rho_c(x)= -\frac{1}{4\pi}(q\hat{K}^{-1}\partial_x\Phi (x)=\frac{1}{2\pi}\frac{1}{3} (\partial_x\Phi_1+\partial_x\Phi_2),
\end{equation}
where matrix $K$ and vector q are given by
\begin{equation}
\mathbf{K}=\left(\begin{array}{cc}1 & 2\\2 & 1 \end{array}\right),
\mathbf{q}=\left(\begin{array}{c}1 \\1 \end{array}\right).
\label{K}
\end{equation}

For interaction matrix $\hat{V}^{e}$, we assume that diagonal matrix elements for intra-mode Coulomb interaction is defined by matrix element $V_1$ and
off-diagonal matrix elements for inter-mode  Coulomb interaction is defined by matrix element $V_2$,
\begin{equation}
\hat{V}^{e}=\left(\begin{array}{cc}V_1 & V_2\\V_2 & V_1 \end{array}\right).
\label{V}
\end{equation}

The commutation relation is
\begin{equation}
\left[ \partial_x \Phi_i(x), \Phi_j(x')\right]=i2\pi K_{ij}\delta(x-x').
\end{equation}
For composite fermion creation operators, we have
\begin{equation}
\Psi_j (x)\propto \exp{-i\Phi_j(x)}
\end{equation}
and
\begin{equation}
\left[\rho_c(x), \Psi_j(x')\right]=\delta(x-x')\Psi_j(x').
\end{equation}

Luttinger liquid action in terms of \textbf{\textit{quasiparticle bosonic fields}} $\varphi$ , defined by $\Phi = {\hat K}\varphi$ is given by
\begin{equation}
S = -\frac{1}{4\pi}\int dt \int dx \left[ (\partial_x \varphi \hat{K}\partial_t\varphi )  +  (\partial_x \varphi\hat{V}^{qp}  \partial_x \varphi)\right],
\end{equation}
where
\begin{equation}
\hat{V}^{qp}=\hat{K}\hat{V}^{e}\hat{K}.
\end{equation}
The two sets of fields are orthogonal:
\begin{equation}
\left[ \partial_x \varphi_i(x), \Phi_j(x')\right]=i2\pi \delta_{ij}\delta(x-x').
\end{equation}

The charge mode $\varphi_c$, and the neutral mode  $\varphi_n$ are expressed as follows:
\begin{equation}
\varphi_{c,n}=\frac{1}{\sqrt{2}}\left( \varphi_1\pm\varphi_2\right),
\end{equation}
where in vector form  $(\varphi_{1},\varphi_{1})^T={\hat W}(\varphi_{c},\varphi_{n})^T$,
opetration $T$ transposing a row vector into a column vector, and the matrix  ${\hat W}$ is given by
\begin{equation}
\mathbf{W}=\left(\begin{array}{cc}1 & 1\\1 & -1 \end{array}\right),
\label{W}
\end{equation}
and the following relations hold
\begin{eqnarray}
\Phi_1 &=\varphi_1 +2\varphi_2 &=\frac{1}{\sqrt{2}}\left( 3\varphi_c-\varphi_n \right)
\label{phi1}\\
\Phi_2 &=\varphi_2 +2\varphi_1&=\frac{1}{\sqrt{2}}\left( 3\varphi_c+\varphi_n \right).
\label{phi2}
\end{eqnarray}
The neutral mode $\varphi_n$ can be interpreted as a difference in the occupation of edge modes corresponding to the first and second $\Lambda$-levels of composite fermions. In the unpolarized phase it coincides with the spin density and we will use spin
index $s$ instead of $n$.

\subsection{Description of edges of $\nu=2/3$ state in terms of separate charge and neutral/spin modes}

In order to discuss the application of voltage to the domain wall system it is convenient to formulate Luttinger liquid action in terms of separate charge and neutral/spin modes. Using matrix $W$ defined by Eq. (\ref{W}) the transformed matrix $K$, given by Eq. (\ref{K}) becomes
\begin{equation}
\mathbf{W^TKW}=\left(\begin{array}{cc}3 & 0\\0 & -1 \end{array}\right).
\label{transformedK}
\end{equation}

For the transformed matrix $\hat{V}^{qp}$,
\begin{equation}
V_{cn}=W^T\hat{V}^{qp}W,
\end{equation}
we obtain
\begin{equation}
V_{cn}=\left(\begin{array}{cc}3v_c & v_{cn}\\v_{cn} & v_s \end{array}\right).
\label{V}
\end{equation}
Here we take off-diagonal terms $v_{cn}=0$ because they are proportional to the difference of velocities of modes $\Phi_1$  and $\Phi_2$.
Then the transformed action is
\begin{equation}
S = -\frac{1}{4\pi}\int dt \int dx \left[ -3\partial_x \varphi_c(\partial_t+v_c\partial_x) \varphi_c   +  \partial_x \varphi_n   (\partial_t+ v_n\partial_x) \varphi_n \right].
\end{equation}
The commutation relations for separated charge and neutral (spin) modes are given by
\begin{eqnarray}
&\left[ \partial_x \varphi_c(x), \varphi_c(x')\right]=& i\frac{2\pi}{3}\delta(x-x')\\
&\left[ \partial_x \varphi_n(x), \varphi_n(x')\right]=&- i2\pi\delta(x-x')
\end{eqnarray}
 To acquire non-zero average charge density and current, density of the charge mode $\varphi_c$ is shifted,
$\varphi_c\rightarrow \varphi_c(x,t) +\bar{\varphi_c}$. A non-zero average appears due to a charge current injection,
\begin{equation}
\bar{\varphi_c}=\frac{e\sqrt{2}}{3\hbar}\left(\frac{x}{v_c}-t\right)V,
\label{shift_1}
\end{equation}
where V is the applied voltage. Then the average current carried by the edge is
\begin{equation}
\bar{j}=-\frac{e}{\sqrt{2}\pi}\partial_t\varphi_c= \frac{e^2}{2\pi\hbar}\frac{2}{3}V
\end{equation}
Shift of the charge mode density is described by an addition to the Luttinger liquid action
\begin{equation}
\Delta S = -\frac{1}{4\pi}\int dt \int dx \sqrt{2} \left(\partial_t+v_c\partial_x\right) \varphi_c ,
\label{shift}
\end{equation}
so that $S(\varphi_c) + \Delta S (\varphi_c)= S(\varphi_c- \bar{\varphi_c})$.

\subsection{The Luttinger liquid action in the presence of spin-polarized and spin-unpolarized phases.}

The Luttinger liquid action for the edge states at $\nu=2/3$ consisting of the two phases, polarized $p$ and unpolarized $u$, is given by
\begin{eqnarray}
&S = -\frac{1}{4\pi}
\int dt \int dx  \left[ (\partial_x \phi_u, \partial_x \phi_p) {\hat{\cal{K}}}^{-1}
\left(\begin{array}{c}\partial_t \phi_u\\
\partial_t \phi_p \end{array}\right) +\right. \nonumber\\
&  \left. (\partial_x \phi_u, \partial_x \phi_p){\hat{\cal{V}}}\left(\begin{array}{c}\partial_x \phi_p\\
\partial_x \phi_p \end{array}\right)\right],
\label{up}
\end{eqnarray}
where 4x4 matrices
 \begin{equation}
{\hat{\cal{K}}}=\left(\begin{array}{cc}
-{\hat{K}} & 0\\
0&
{\hat{K}} \end{array}\right),
\end{equation}
with matrix ${\hat{K}}$ defined by Eq. (\ref{K}) and
\begin{equation}
{\hat{\cal{V}}}=\left(\begin{array}{cc}
-{\hat{V}^{e}} & 0\\
0&
{\hat{V}^{e}} \end{array}\right),
\end{equation}
with matrix $\hat{V}^{e}$ is defined by Eq. (\ref{V}). For convenience, in order to keep the form of relations for the current as described above for both p and u phases, we reverse the sign of the quasiparticle field $\varphi_u\rightarrow -\varphi_u$, so that $\Phi_u=K\varphi_u$.

\subsection{Tunneling and charge currents}

The tunneling between polarized and unpolarized phases for modes carried by composite fermions of the same spin polarization is described by the tunnel Hamiltonian
\begin{eqnarray}
&{\cal H}_T=-\bar{t}\cos{(\Phi_{p1}(0,t)- \Phi_{u1}(0,t))}\nonumber \\
&=-\bar{t}\cos{\frac{1}{\sqrt{2}}(3\varphi_{pc}(0,t)- \varphi_{pn}(0,t)-3\varphi_{uc}(0,t)+ \varphi_{us}(0,t))},
\end{eqnarray}
where we re-labeled the neutral mode in the unpolarized phase as a spin mode describing the difference in spin density between modes.

The tunneling charge current is given by a shift in $\varphi_{pc}$ due to the applied voltage $V$ described by Eqs. (\ref{shift_1}, \ref{shift}):
 \begin{equation}
J_T=-\partial_t \hat{Q_P}(t)=i\left[\int dx \rho_{pc}, {\cal H}_T\right]= -\bar{t}\sin{(\Phi_p(0,t)- \Phi_u(0,t))},
\end{equation}
where $\rho_{pc}=\frac{1}{\sqrt{2}\pi}\partial_x\varphi_{pc}$ is the charge density.

\subsection{Tunneling in the model of zero length hDW}

In the strong coupling limit the tunneling current can be found by imposing conditions. A zero-length hDW is the limit $x_1=x_2=0$ of the model of the hDW shown in Fig.~3a of the main text.
In order to formulate the boundary conditions, we use quasiparticle modes
 described by Eqs.(\ref{phi1},\ref{phi2}) for polarized and unpolarized liquid. Imposing ${\cal H}_T=-\bar{t}\cos{(\Phi_{p1}(0,t)- \Phi_{u1}(0,t))}$ at $\bar{t}\rightarrow \infty$ as a boundary condition that leads to a jump in the tunneling mode and a continuity in the orthogonal, non-tunneling mode, we obtain:
 \begin{eqnarray}
&\Phi_{p1}(-0)- {\Phi}_{u1}(+0)=-(\Phi_{p1}(+0)- \Phi_{p1}(-0),\\
&\Phi_{p1}(-0)+\Phi_{u1}(+0)=-\Phi_{p1}(+0)+ \Phi_{u1}(-0).
\label{bound12}
\end{eqnarray}

Using the expressions for quasiparticle modes (\ref{phi1}), we have
\begin{eqnarray}
&3\varphi^{\rightarrow}_{pc}(-0)-\varphi^{\leftarrow}_{pn}(+0)=3\varphi^{\leftarrow}_{uc}(-0)-\varphi^{\rightarrow}_{us}(+0),\\
&3\varphi^{\leftarrow}_{uc}(+0)-\varphi^{\rightarrow}_{us}(-0)=3\varphi^{\rightarrow}_{pc}(+0)-\varphi^{\leftarrow}_{pn}(-0).
\end{eqnarray}

We obtain two more equations defining boundary conditions imposing them for the two modes orthogonal to (\ref{phi1}). These modes are given by
 \begin{eqnarray}
&\phi_{u2}(x)=\frac{1}{\sqrt{2}}(\varphi_{uc}(x)+\varphi_{us}(x)),\\
&\phi_{p2}(x)=\frac{1}{\sqrt{2}}(\varphi_{pc}(x)+\varphi_{pn}(x)).
\label{barphi2}
 \end{eqnarray}
The boundary conditions for these modes are
 \begin{eqnarray}
&\varphi_{uc}(-0)-\varphi_{us}(+0))= \varphi_{uc}(+0)-\varphi_{us}(-0)),\\
&\varphi_{pc}(-0)-\varphi_{pn}(+0))= \varphi_{pc}(+0)-\varphi_{pn}(-0)).
\label{bound34}
\end{eqnarray}

Using explicit expressions for the quasiparticle modes, we obtain that Eqs.( \ref{bound12},\ref{bound34}) result in the following four equations defning the outgoing fields via incoming fields:
\begin{eqnarray}
&4\varphi^{\rightarrow}_{pc}(+0)=3\phi^{\leftarrow}_{uc}(+0)-\varphi^{\rightarrow}_{us}(-0)+\varphi^{\rightarrow}_{pc}(-0)+\varphi^{\leftarrow}_{pn}(+0),\nonumber\\
&4\varphi^{\leftarrow}_{pn}(-0)=3\varphi^{\rightarrow}_{pc}(-0)+3\varphi^{\leftarrow}_{pn}(+0)-3\varphi^{\leftarrow}_{uc}(+0)+\varphi^{\rightarrow}_{us}(-0),\nonumber\\
&4\varphi^{\leftarrow}_{uc}(-0)=3\varphi^{\rightarrow}_{pc}(-0)-\varphi^{\leftarrow}_{pn}(+0)+\varphi^{in}_{uc}(+0)+\varphi^{\rightarrow}_{us}(-0),\nonumber\\
&4\varphi^{\rightarrow}_{us}(+0)=3\varphi^{\leftarrow}_{uc}(+0)+3\varphi^{\rightarrow}_{us}(-0)-3\varphi^{\rightarrow}_{pc}(-0)+\varphi^{\leftarrow}_{pn}(+0).
\label{inout}
\end{eqnarray}

The current injected into the polarized phase due to the applied voltage $V$ shifts only the incoming field
\begin{equation}
\varphi^{\rightarrow}_{pc}(-0)\rightarrow \varphi^{\rightarrow}_{pc}(-0)-\frac{\sqrt{2}}{3}eVt/\hbar.
\label{shiftdens}
\end{equation}
Using Eq.(\ref{inout}), we obtain that this change leads to the following changes in outgoing fields:
\begin{eqnarray}
&\varphi^{\rightarrow}_{pc}(+0)\rightarrow \varphi^{\rightarrow}_{pc}(+0) -\frac{1}{4}\frac{\sqrt{2}}{3}eVt/\hbar,\nonumber\\
&\varphi^{\leftarrow}_{pn}(-0)\rightarrow \varphi^{\leftarrow}_{pn}(-0)-\frac{3}{4}\frac{\sqrt{2}}{3}eVt/\hbar,\nonumber\\
&\varphi^{\leftarrow}_{uc}(-0)\rightarrow \varphi^{\leftarrow}_{uc}(-0)-\frac{3}{4}\frac{\sqrt{2}}{3}eVt/\hbar,\nonumber\\
&\varphi^{\rightarrow}_{us}(+0)\rightarrow \varphi^{\rightarrow}_{us}(+0)+\frac{3}{4}\frac{\sqrt{2}}{3}eVt/\hbar.
\label{shiftout}
\end{eqnarray}

\begin{figure}
\vspace{-1cm}
\centering
\includegraphics[width=0.6\textwidth]{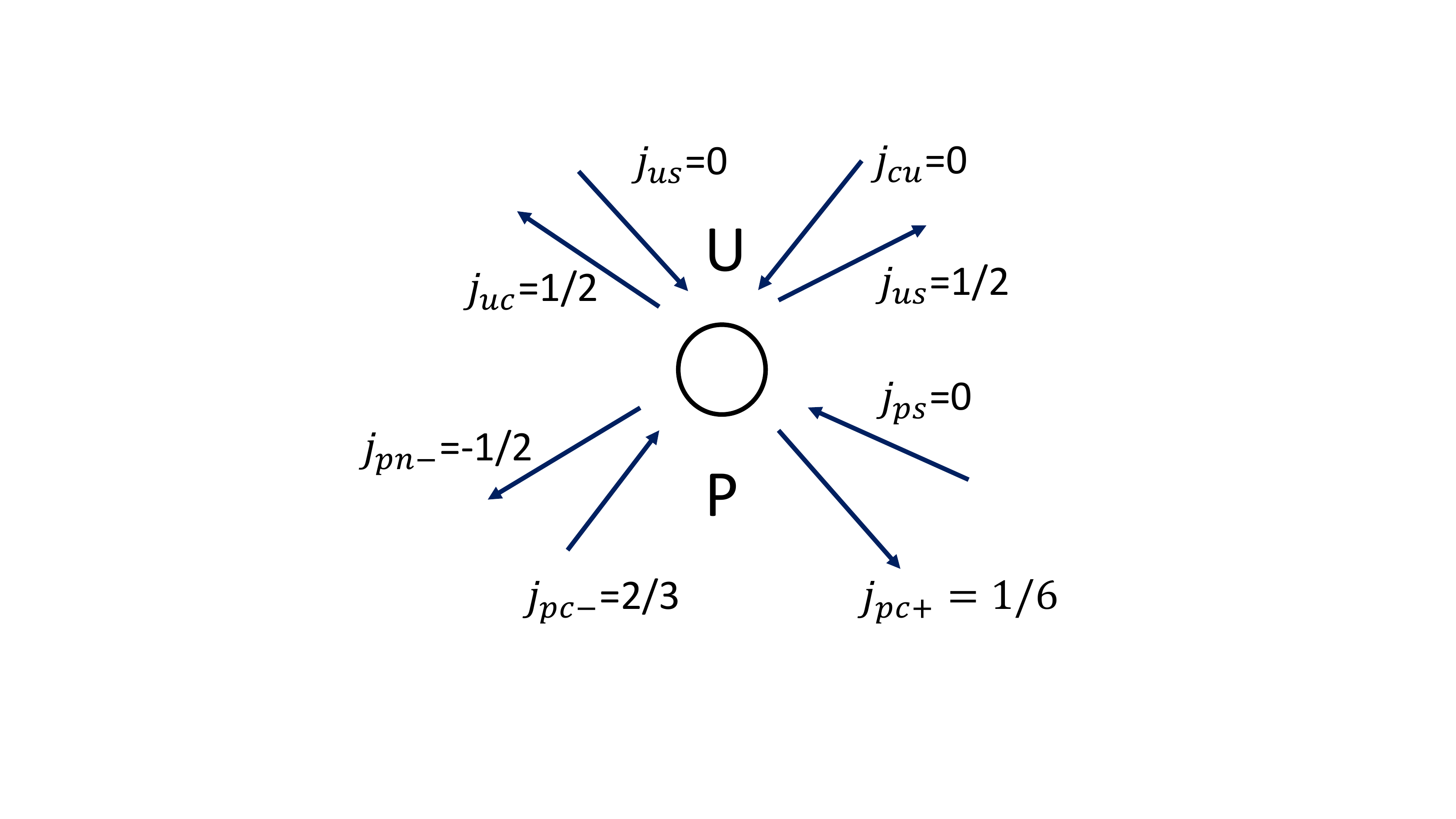},
\vspace{-1cm}
\caption{ Incoming and outgoing currents in a single point contact in units of $\sigma_0V$. } \label{fig1}
\end{figure}

The average incoming or outgoing current due to any of the quasiparticle modes $\varphi$ (or $\eta$) is given by the equation
\begin{equation}
j=-e q\sqrt{2} \partial_t\varphi,
\end{equation}
where for our choice of signs in action Eq. (\ref{up}) charges $q_p=1=-q_u$, $q_p$ is the quasiparticle charge in the polarized phase and
$q_u$ is the quasiparticle charge in the unpolarized phase.
Describing the currents, we will use indices $p$ and $u$ for the currents on the polarized and unpolarized side, correspondingly. Upper indices $in$, $out$ correspond to the incoming and outgoing currents, lower indices $c$, $n$ and $s$
correspond to charge, neutral and spin currents. On the polarized side,
charge and spin currents coinside; on the unpolarized side, neutral and spin currents coinside. We will assume first that the externally induced incoming charge current comes from the contact with potential $V$ to the polarized liquid, and the current flows into the grounded ($V=0$) contact to the unpolarized liquid.
Then the incomig current due to (\ref{shiftdens}) is
\begin{equation}
j^{in}_{pc}=\sigma_0\frac{2}{3}V,
\end{equation}
where $\sigma_0=e^2/h$ is the conductance quantum, and the outgoing currents are given by
\begin{eqnarray}
&j^{out}_{pc}=\sigma_0\frac{1}{6}V\\
&j^{out}_{pn}=-\sigma_0\frac{1}{2}V\\
&j^{out}_{uc}=+\sigma_0\frac{1}{2}V\\
&j^{out}_{us}=\sigma_0 \frac{1}{2}V,
\end{eqnarray}
where signs of $j^{out}_{pn}$ and $^{out}_{uc}$  are changed compared to Eq.(\ref{shiftout}) as we transition to counting signs from the right-moving direction, see Fig\ref{fig1}.

Using the voltage-induced shift of the incoming and outgoing bosonic fields, ${\varphi}_{pc}$,   ${\varphi}_{ps}$,  ${\varphi}_{uc}$ and  ${\varphi}_{us}$, given by Eqs (\ref{shiftdens}), (\ref{shiftout}),
we can analyze the shift of quasiparticle edge state fields  ${\varphi}_{p(1,2)}=\frac{1}{\sqrt{2}}({\varphi}_{pc}\pm{\varphi}_{pn})$   and  ${\varphi}_{u(1,2)}=\frac{1}{\sqrt{2}}({\varphi}_{uc}\pm{\varphi}_{us})$, that describes the distribution of current over the two quasiparticle modes.
The calculation shows that the resulting currents associated with these modes are given by
\begin{eqnarray}
&j_{p1}=\sigma_0V\left[ \frac{7}{12}\theta(-x)+\frac{1}{12}\right]\\
&j_{p2}=\sigma_0\frac{1}{12}V\\
&j_{u1}=-\sigma_0V\left[-\frac{1}{4}\theta(-x)+\frac{1}{4}\theta(x)\right]\\
&j_{u2}=-\sigma_0\frac{1}{4}V,
\end{eqnarray}
where $\theta$-function $\theta(x)=1$ at $x>0$ and  $\theta(x)=0$ at $x<0$ is used to describe incoming and outgoing modes in a single equation.
We observe that it follows from these equations that tunneling occurs only between   ${\bar\varphi}_{p1}$ and
${\bar\varphi}_{u1}$ edges,
\begin{equation}
J_T=\sigma_0V(\frac{2}{3}-\frac{1}{6})=\frac{1}{2}\sigma_0V,
\end{equation}
while states ${\bar\varphi}_{p2}$ and ${\bar\varphi}_{u2}$
flow, correspondingly, in polarized and unpolarized region. Modes  ${\bar\varphi}_{p1}$ and
${\bar\varphi}_{u1}$ include modes with the same spin up, while modes  ${\bar\varphi}_{p2}$ and ${\bar\varphi}_{u2}$ carry the opposite spin. This is a consequence of our tunnel Hamiltonian allowing only transmission of like spins. Generalization of the model permitting
tunneling with a spin flip, e.g,. induced by interaction with nuclear spins,  makes possible some tunneling processes between  ${\bar\varphi}_{p2}$ and ${\bar\varphi}_{u2}$ modes.

\subsection{Domain wall of finite length with scattering at the ends}

We now combine two point scatterers at points $x_1$ and $x_2$ separated a ballistic domain wall of length $L=x_2-x_1$.  In the experimental setting, these scatterers are the tri-junctions between edge states and the domain wall.
\begin{figure}
\vspace{-1cm}
\centering
\includegraphics[width=0.6\textwidth]{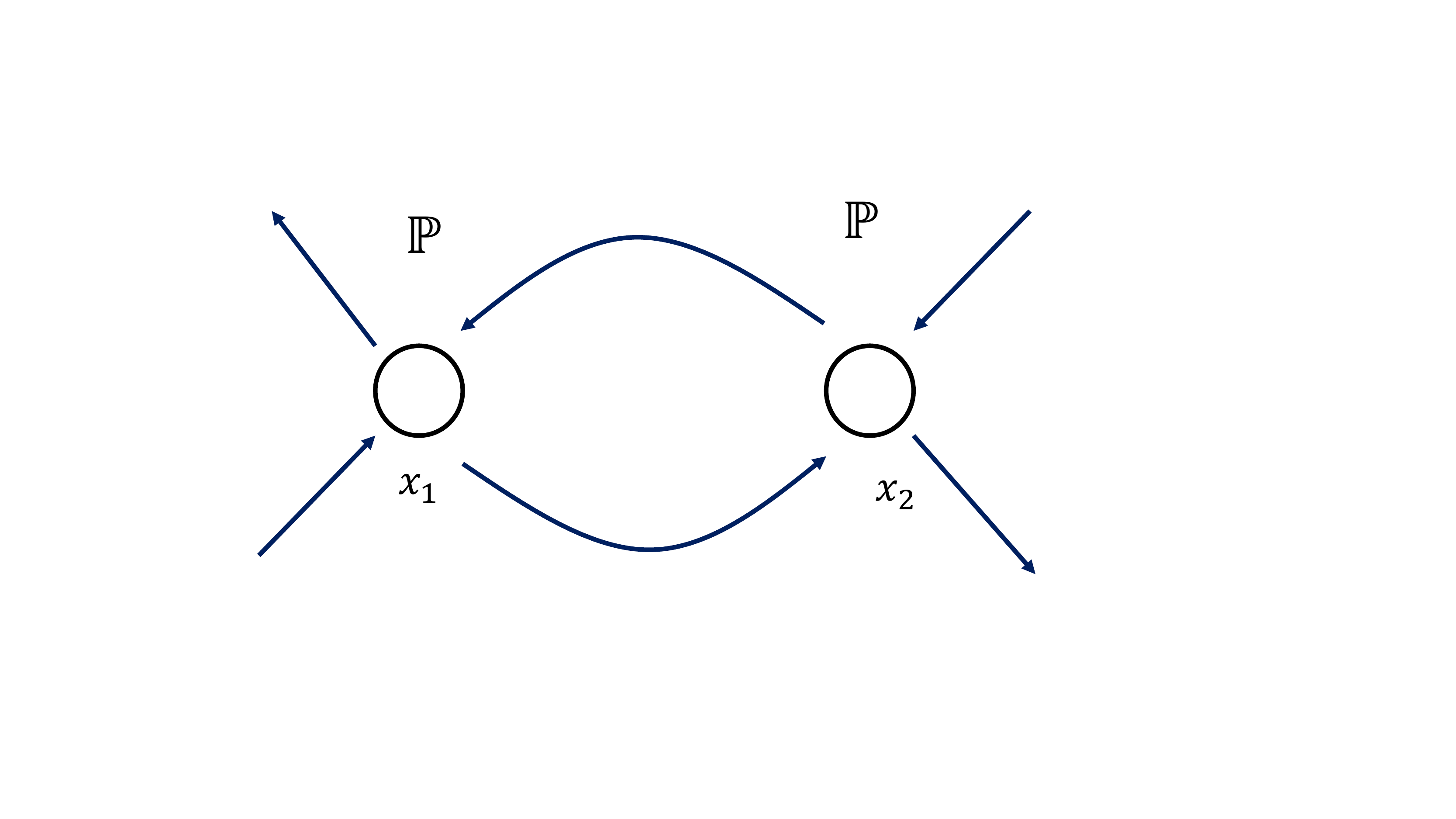},
\vspace{-1cm}
\caption{ Incoming and outgoing modes in the case of two junctions  positioned at $x_1$ and $x_2$, desribing the total transfer matrix {\cal T}. Each junction is described by the propagation matrix ${\cal P}$ given by Eq. (\ref{calP}) for the zero-lenth hdW. } \label{fig2}
\end{figure}

Each of the scatterers is described by Eq. (\ref{inout}). To calculate the transmission through two junctions, it is convenient to
present the connection  (\ref{inout}) between 4-vector of  incoming  modes $\varphi^{in}=(\varphi^{\rightarrow}_{pc}(x-0),\varphi^{\rightarrow}_{us}(x-0),\varphi^{\leftarrow}_{uc}(x+0),
\varphi^{\leftarrow}_{pn}(x+0))^T$ and  4-vector of  outgoing modes $\varphi^{out}=(\varphi^{\rightarrow}_{pc}(x+0),\varphi^{\rightarrow}_{us}(x+0),\varphi^{\leftarrow}_{uc}(x-0),\varphi^{\leftarrow}_{pn}(x-0))^T$ in a matrix form:
\begin{equation}
\varphi^{out}= {\cal P}\varphi^{in},
\end{equation}
where 4x4 matrix  ${\cal P}$ is given by
\begin{equation}
 {\cal P}=\left(\begin{array}{cc}P_{++} & P_{+-} \\P_{-+}  & P_{--} \end{array}\right),
\label{calP}
\end{equation}
and 2x2 matrices $P_{ij}$,$ i,j=\pm$ are defined by
\begin{equation}
 P_{++}=P_{--}=\frac{1}{4}\left(\begin{array}{cc}1 & -1 \\-3 &3\end{array}\right),
\label{Pii}
\end{equation}
\begin{equation}
 P_{+-}=P_{-+}=\frac{1}{4}\left(\begin{array}{cc}3& 1 \\3 &1\end{array}\right).
\label{Pij}
\end{equation}
The matrix $ P_{++}$ describes propagation of chiral modes through a single zero-length scatterer, and the matrix  $ P_{+-}$ describes the reflection of chiral modes.
The total 4x4 transfer matrix ${\cal T}$ connects incoming and outgoing modes of two scatterers, via
 \begin{equation}
\begin{pmatrix}
\Delta_V\varphi^{\rightarrow}_{pc}(x_2+0)\\[\jot]
\Delta_V\varphi^{\rightarrow}_{us}(x_2+0)\\[\jot]
\Delta_V\varphi^{\leftarrow}_{uc}(x_1-0)\\[\jot]
\Delta_V\varphi^{\leftarrow}_{pn}(x_1-0)\end{pmatrix}
= {\cal T}\begin{pmatrix}
\Delta_V\varphi^{\rightarrow}_{pc}(x_1-0)\\[\jot]
\Delta_V\varphi^{\rightarrow}_{us}(x_1-0)\\[\jot]
\Delta_V\varphi^{\leftarrow}_{uc}(x_2+0)\\[\jot]
\Delta_V\varphi^{\leftarrow}_{pn}(x_2+0)\end{pmatrix},
\end{equation}
where $x_1$ and $x_2$ are positions of the scatterers and $\Delta_V$ denotes a voltage-dependent shift of the corresponding field.

Matrix ${\cal T}$ is defined by propagation of modes through the scatterers, and potentially by multiple reflections of modes between them. However, as a result of identity
\begin{equation}
 P_{++}P_{-+}=0,
\label{disentangle}
\end{equation}
the total transfer matrix for two-junction system is defined only by a single act of reflection, $T_{-+}=P_{-+}$ or by a single propagation through both scatterers, $T_{++}=P_{++}P_{++}=P_{++}$, while any contribution from processes involving subsequent propagations and reflections from scatterers vanish due to disentanglement property of chiral channels described by Eq. (\ref{disentangle}).

\subsection{Account for tunneling between the same spin modes in the polarized region}

We now consider whether tunneling between the same spin modes in the polarized region between two zero length scatterers will change transmission through the domain wall. The boundary condition introduced by the tunneling Hamiltonian corresponding to this process
\begin{equation}
{\cal H}_T=-\bar{t}_{p}\cos{(\Phi_{p1}(x)- \Phi_{p2}(x))}=-\bar{t}_p\cos{\frac{1}{\sqrt{2}}\varphi_{pn}(x)},
\end{equation}
where $x_1<x<x_2$, in the strong coupling limit $\bar{t}_{p}\rightarrow\infty$ is
\begin{equation}
\varphi_{pn}(x-0)=-\varphi_{pn}(x+0).
\end{equation}
Taking into account this process amounts to changing one of the $\cal{P}$ matrices in the model with two scatterers
\begin{equation}
{\cal P}_1= \left(\begin{array}{cccc}1 & 0&0&0\\0 & 1&0&0\\0&0&1&0\\0&0&0&-1 \end{array}\right)\cal{P}.
\end{equation}
We then observe that $ P_{1++}= P_{++}$, $P_{1-+}=P_{-+}$, as described by Eqs.(\ref{Pii}),(\ref{Pij}) and
\begin{equation}
P_{1--}=\frac{1}{4}\left(\begin{array}{cc}1 & 1 \\-3 &-3\end{array}\right),
\label{P1}
\end{equation}
\begin{equation}
 P_{1+-}=\frac{1}{4}\left(\begin{array}{cc}3& -1 \\3 &-1\end{array}\right).
\label{P1r}
\end{equation}
Therefore, disentangling relations
\begin{eqnarray}
& P_{1++}P_{1+-}=0\\
& P_{1--}P_{1-+}=0\\
& P_{1-+}P_{1++}=0\\
& P_{1+-}P_{1--}=0
\end{eqnarray}
hold, and no contributions from processes with consequent propagation and reflection occur into total transfer matrix in the presence of
$P1\rightarrow P2$ scattering. The total transfer matrix ${\cal T}_c$ for this case is also defined by
\begin{eqnarray}
& T_{c++}=P_{++}\\
&  T_{c+-}=P_{+-}\\
& T_{c-+}=P_{-+}\\
& T_{c--}=P_{--}
\end{eqnarray}
That means, ${\cal T}_c={\cal P}$ despite scattering in the same spin channel in the polarized region.
That leads us to conclusion that localization and backscattering by the domain wall that leads to length-dependent resistance requires spin flips or inelastic processes.

\subsection{Charge, neutral and spin currents}
\begin{figure}
\vspace{-1cm}
\centering
\includegraphics[width=0.8\textwidth]{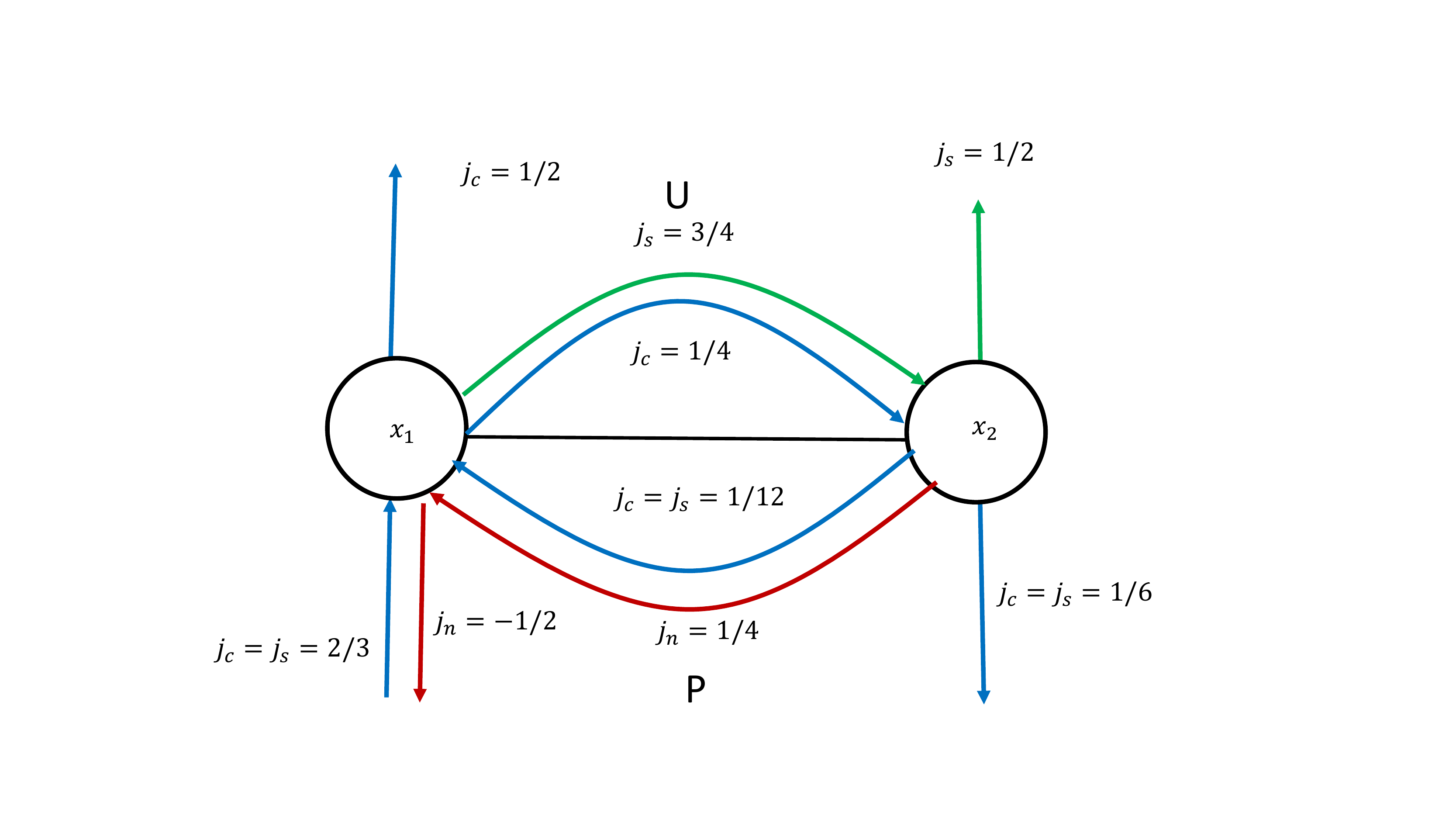},
\vspace{-1cm}
\caption{The current flow for current $j_c=2/3$ injected from the polarized into the unpolarized region. Values of currents are shown in units of $\sigma_0 V$. Charge currents in the polarized and unpolarized regions are shown in blue. In the polarized region, the charge current and the spin current coinside. The neutral current in polarized region, describing the difference of currrents carried by two quasiparticle modes, is shown in red. Spin currents in the unpolarized region, corresponding to the difference of currents carried by spin up and spin down modes in there, are shown in green.} \label{currents}
\end{figure}

We now summarize our results, see Fig.~\ref{currents}. We use
indices $1$ and $2$  to describe the incoming (index $in$) and outgoing (index $out$) currents flowing on the sample edges through the  scatterers at $x_1$ and $x_2$ , where the
edge channels at the boundary of the sample are coupled to the channels propagating inside the domain wall. These scatterers are triple junctions that simultaneously couple edges at the boundary in the polarized phase couple to edges in the unpolarized phase. The currents (index $DW$) inside the domain wall flow from left to right (+sign) and right to left (-sign).
  We then have
\begin{eqnarray}
&j_{pc}^{in,1}&= \frac{2}{3}\sigma_0 V, \hspace{1cm}
j_{uc}^{out,1}= \frac{1}{2}\sigma_0 V \nonumber\\
&j_{us}^{in,1}&= 0\hspace{2cm}
j_{pn}^{out,1}= -\frac{1}{2}\sigma_0 V\nonumber\\
&j_{pn}^{in,2}&= 0\hspace{2cm}
j_{pc}^{out,2}= \frac{1}{6}\sigma_0 V\nonumber\\
&j_{us}^{out,2}&= -\frac{1}{2}\sigma_0 V\hspace{1cm}
j_{uc}^{in,2}= 0\nonumber\\
&j_{uc}^{DW}&= \frac{1}{4}\sigma_0 V
\hspace{1.4cm}
j_{us}^{DW}= \frac{3}{4}\sigma_0 V\nonumber\\
&j_{pc}^{DW}&= -\frac{1}{12}\sigma_0 V
\hspace{1cm}
j_{pn}^{DW}= -\frac{1}{4}\sigma_0 Vt
\label{currents_theory_p}
\end{eqnarray}
It is easy to see that at each scatterer, $x_1$ and $x_2$, the charge current is conserved, and so are the spin currents in the unpolarized liquid, and the neutral currents in the polarized liquid. Also, spin projection of incoming electrons is equal to spin projection of outgoing electrons at junctions $x_1$ and $x_2$.

The case of incoming charge current from the unpolarized liquid flowing into polarized liquid, is described by similar equations. The
the result for charge currents then is given by Eq.(\ref{currents_theory_p}) with the permutation
$p\leftrightarrow u$. For spin and neutral currents, the resultant equations are given by Eq.(\ref{currents_theory_p}) with simultaneous permutation $p,n \leftrightarrow u,s$. Similarly, there are conservation laws for charge currents, spin currents in the unpolarized phase, neutral currents in the polarized phase at the $x_1$ and $x_2$ scatterers on the edge of the sample. The domain wall in both cases, for the current injected from the \textit{p}-phase into the \textit{u}-phase, and for the current injected from the \textit{u}-phase into the \textit{p}-phase, can be described by the ballistic conductance equal to $1/3e^2/h$.


\end{document}